\documentclass[a4paper,longbibliography,aps,prb,preprintnumbers,amsmath,amssymb,twocolumn,superscriptaddress]{revtex4-1} 

\usepackage{amsmath}
\usepackage{amssymb}
\usepackage{verbatim}
\usepackage{graphicx}
\usepackage{color}
\usepackage{xspace}
\usepackage{bm}

\begin{document}

\title{Two-dimensional excitons from twisted light and the fate of the photon's orbital angular momentum}

\author{Tobias Gra{\ss}}
\affiliation{ICFO-Institut de Ciencies Fotoniques, The Barcelona Institute of Science and Technology, 08860 Castelldefels (Barcelona), Spain}
 \author{Utso Bhattacharya}
 \affiliation{ICFO-Institut de Ciencies Fotoniques, The Barcelona Institute of Science and Technology, 08860 Castelldefels (Barcelona), Spain}
 \author{Julia Sell}
 \affiliation{Joint Quantum Institute, NIST and University of Maryland, College Park, Maryland, 20742, USA}
 \author{Mohammad Hafezi}
 \affiliation{Joint Quantum Institute, NIST and University of Maryland, College Park, Maryland, 20742, USA}
 \affiliation{Department of Electrical Engineering and Institute for Research in Electronics and Applied Physics, University of Maryland, College Park, MD 20742, USA}

\begin{abstract}
    As the bound state of two oppositely charged particles, excitons emerge from optically excited semiconductors as the electronic analogue of a hydrogen atom. In the two-dimensional (2D) case, realized either in quantum well systems or truly 2D materials such as transition metal dichalcogenides, the relative motion of an exciton is described by two quantum numbers: the principal quantum number $n$, and a quantum number $j$ for the angular momentum along the perpendicular axis. Conservation of angular momentum demands that only the $j=0$ states of the excitons are optically active in a system illuminated by plane waves. Here we consider the case for spatially structured light sources, specifically for twisted light beams with non-zero orbital angular momentum per photon.  Under the so-called dipole approximation where the spatial  variations of the light source occur on length scales much larger than the size of the semiconductor's unit cell, we show that the photon (linear and/or angular) momentum is coupled to the center-of-mass (linear and/or angular) momentum of the exciton. Our study establishes that the selection rule for the internal states of the exciton, and thus the exciton spectrum, is independent from the spatial structure of the light source.
\end{abstract}

\maketitle

\section{Introduction}

Excitons are the bound states formed by an electron-hole pair in a semiconductor crystal \cite{schaefer-book,haugkoch}, and as such, they are close analogues of the hydrogen atom. Excitons manifest themselves as optical absorption or emission lines within the band gap of the material. In the theoretical treatment of exciton formation, light-matter interaction is most often described within the dipole approximation. This approximation disregards the spatial structure of the light field, which is justified by the tiny length scale on which the de Broglie waves vary, as compared to the wavelength of the light \cite{cohentannoudji}.

The dipole approximation gives rise to optical selection rules related to the conservation of angular momentum during an optical transition. Within the dipole approximation, light may carry only one quantum of angular momentum per photon, realized through the circular polarization of the light; therefore, optical transitions can change angular momentum quantum numbers of the matter only by one unit. In atoms, this rule selects the $s$-to-$p$ or $p$-to-$d$ transitions; in semiconductors, the selection rule affects the orbitals of the bands in an analogous way. As a consequence of the dipole approximation, excitons created from such a dipole transition do not carry angular momentum; that is, the respective quantum number $j$ is zero. Moreover, as established by the Elliott formula \cite{elliott57}, the transition amplitude quickly decays with the principal quantum $n$ (as $(n+1/2)^{-3}$ in 2D), such that the exciton spectrum is strongly dominated by transitions into the 1$s$ state, i.e. the state corresponding to the hydrogenic ground state. A common technique to allow optical access to the $p$ exciton series in semiconductors is nonlinear, two-photon spectroscopy \cite{berkelbach15,wang15}. 

It is clear that effects beyond the dipole approximation can modify the selection rule of the excitons. In particular, the dipole approximation disregards the possible spatial structure of the light beam, which in the case of twisted light results in a well-defined orbital angular moment (OAM) per photon \cite{allen92,yao11,torres2011twisted}. The photon OAM essentially adds another tunable degree of freedom for tailoring light-matter interaction. It has been proposed to use this new degree of freedom for generating a topological band structure by breaking time-reversal symmetry \cite{bhattacharya21}, for pumping electrons in a magnetic field through the Landau level  \cite{grass18,fujita19,cao21}, or for producing topological defects such as vortices or skyrmions \cite{fujita17,fujita172,cian20,kim21}.
A striking demonstration of how optical selection rules are modified by photon OAM has been achieved in an experiment with trapped ions \cite{schmiegelow16}, showing a dipole-forbidden atomic $s$-to-$d$ transition in the presence of a twisted light field. Given the analogy between an exciton and a hydrogen atom, it might be expected that twisted light can generate transitions into dark excitonic levels, where the photonic OAM is absorbed in the internal degree of freedom. This indeed has theoretically been suggested for the case of Rydberg excitons \cite{konzelmann19}. On the other hand, there have also been experiments with atomic and polaritonic condensates in which twisted light has led to the formation of vortices \cite{andersen2006,kwon19}. This indeed would suggest that the orbital angular momentum of the photon is absorbed by the center-of-mass (COM) degree of freedom of the exciton, rather than by the relative motion of electron and hole. Also, in the strong drive limit and absence of Coulomb binding, OAM of light can lead to Floquet vortex creation, but it is an open question how strong drive and exciton formation compete with each other.

To better understand the fate of the photonic OAM in excitonic transitions, the present paper studies the case of a single exciton in a two-band semiconductor model in 2D in the presence of a twisted light source. We extend theoretical studies of band-to-band transitions in semiconductors or graphene with twisted light, presented in Refs. \onlinecite{quinteiro09,quinteiro10,farias13}, to the case where Coulomb interactions give rise to exciton formation. Our analysis demonstrates that, under the assumption that the spatial variation of the light occurs on a length scale much larger than the size of the unit cell of the semiconductor crystal, transitions into excitonic levels $j\neq 0$ remain completely forbidden even in the presence of twisted light. Instead, the structure of the light field selects the COM degree of freedom of the excitons. Since it is the relative motion that essentially determines the energy of an exciton, it follows that the twist of the light source does not modify the excitonic spectrum. In this context, we also note that small shifts of the spectrum are possible if the COM dispersion of the exciton is taken into account. This indeed has been observed in a recent experiment with excitons in a Dirac material, which found a blueshift of the exciton lines for sufficiently large values of photon OAM \cite{simbulan20}.

 Our paper is organized in the following way: In Sec. II, we develop the general analytical formalism to describe exciton transitions in structured light beam. In Sec. III, we specifically address the case of a Bessel beam. To evaluate this case, we make use of the rotational symmetry of the beam which makes an explicit numerical treatment feasible. With this we are able to show, for a finite system size, that the $s$ states are optically bright, in quantitative good agreement with the 2D Elliot formula, independent from the choice of the photon OAM. Our numerical calculation also confirms  that the COM momentum of the exciton is peaked at the linear momentum of the photon. 

\section{General Analytical Model}
\subsection{Light-matter coupling}
We consider a 2D semiconductor with Bloch bands $\lambda$ and wave vector ${\bf k}$, described by Bloch functions $\varphi_{\lambda,{\bf k}}({\bf r}) = \frac{1}{\sqrt{S}} e^{i {\bf k}\cdot {\bf r}} u_{\lambda,{\bf k}}({\bf r})$, where $u_{\lambda,{\bf k}}({\bf r}+{\bf R_i}) = u_{\lambda,{\bf k}}({\bf r})$ with ${\bf R}_i$ a lattice vector. 
In this basis, the crystal Hamiltonian reads $H_0 = \sum_{\lambda,{\bf k}} \epsilon_{\lambda,\bf k} c^{\dagger}_{\lambda,{\bf k}} c_{\lambda,{\bf k}}$,
with $ c_{\lambda,{\bf k}}$  ($c^\dagger_{\lambda,{\bf k}}$) being the annihilation (creation) operators, and $\epsilon_{\lambda,\bf k}$ the dispersion. We assume a light field given by a vector potential ${\bf A}({\bf r})=A({\bf r}) \cdot {\bf e}$ in the Coulomb gauge, such that the light-matter Hamiltonian is given by:
\begin{align}
 H_{\rm LM} = \sum_{\lambda',\lambda} \sum_{{\bf k}',{\bf k}} \frac{i e \hbar}{M} \langle \lambda', {\bf k}' | {\bf A}({\bf r}) \cdot \nabla_{\bf r} | \lambda, {\bf k} \rangle c^\dagger_{\lambda',{\bf k}'} c_{\lambda, {\bf k}}.
\end{align}
We are only interested in the matrix element $h^{\lambda',\lambda}_{{\bf k}',{\bf k}} = \langle \lambda', {\bf k}' | {\bf A}({\bf r}) \cdot \nabla_{\bf r} | \lambda, {\bf k} \rangle = \int d^2r \varphi^*_{\lambda',{\bf k}'}({\bf r}) A({\bf r}) {\bf e}\cdot\nabla_{\bf r} \varphi_{\lambda,{\bf k}}({\bf r})$ with $\lambda={\rm c}$ and $\lambda'={\rm v}$, i.e. transitions amplitudes between conduction and valence band. Taking into account the orthonormality of the bands, the derivative operator has to act onto the lattice-periodic function $u_{\lambda,{\bf k}}$ to yield non-zero contributions. Explicitly, we have
\begin{align}
 h^{\rm v,c}_{{\bf k}',{\bf k}} = \frac{1}{S} \int d^2r A({\bf r}) e^{i({\bf k}-{\bf k}')\cdot {\bf r}} u^*_{{\rm v}, {\bf k}'}({\bf r}) {\bf e}\cdot\nabla_{\bf r} u_{{\rm c},{\bf k}}({\bf r}).
 \label{hvc}
\end{align}
Here, $S$ is the size of the system.

At this stage, we make the approximation {\bf (A1)}: the vector potential and the exponential do not vary within a unit cell. By keeping variations beyond the scale of a unit cell, this approximation is less restrictive than the dipole approximation which would fully ignore the spatial structure of the light. Yet without considering a particular choice of vector potential, the spatial variations of the beam is generally limited by a length scale on the order of the wavelength of the light. The same length scale also determines the variation of the exponential $e^{i({\bf k}-{\bf k}')\cdot{\bf r}}$, since it will turn out \textit{a posteriori} that ${\bf k}-{\bf k}'$ is determined through the photon momentum. Since the optical wavelength is usually several orders of magnitude larger than the size of unit cell, the approximation is totally valid in the usual cases, but might not hold in some special cases, e.g. of Moir\'{e} lattices with enlarged unit cells \cite{seyler19,tran19,jin19,alexeev19}. Applying (A1) to Eq.~(\ref{hvc}), we write:
\begin{align}
 h^{\rm v,c}_{{\bf k}',{\bf k}} &= \frac{1}{N}_{\rm sites} \sum_{{\bf R}_i} A({\bf R}_i) e^{i({\bf k}-{\bf k}')\cdot {\bf R}_i} \times \nonumber \\ &
 \times \frac{1}{S_{\rm cell}} {\bf e}\cdot \left[ \int_{\rm cell} d^2r u^*_{{\rm v},{\bf k}'}({\bf r}) \nabla_{\bf r} u_{{\rm c},{\bf k}}({\bf r}) \right] 
 \equiv \nonumber \\ & \equiv
 A_{\bm \kappa} \times {\bf e}\cdot {\bf p}_{{\bf k}',{\bf k}}^{\rm vc},
 \label{eqn:hvc}
 \end{align}
where ${\bm \kappa}={\bf k}-{\bf k}'$ and $A_{\bm \kappa}= \frac{1}{N}_{\rm sites} \sum_{{\bf R}_i} A({\bf R}_i) e^{i {\bm \kappa} \cdot {\bf R}_i}$ the Fourier transform of the vector potential. The dipole moment between the ${\bf k}'$ state in the valence band and the ${\bf k}$ state in the conduction band is denoted by ${\bf p}_{{\bf k}',{\bf k}}^{\rm vc}$.

We proceed by making a second approximation {\bf(A2)}: the dipole moment depends only weakly on the wave vectors ${\bf k}$ and ${\bf k}'$. Indeed, the most radical implementation of this approximation in which the dipole moment is set to a constant ${\bf p}^{\rm vc}_0$ is commonly used in the literature, cf. Ref. \onlinecite{haugkoch}. To be less restrictive, we argue that ${\bf p}_{{\bf k}',{\bf k}}^{\rm vc}$ may depend on ${\bf k}+{\bf k}'$ (which can take relatively large values), whereas the dependence on ${\bf k}-{\bf k}'$ (which remains small since it is equivalent to the photon momentum) is neglible. To this end we introduce the quantity ${\bf K}= \frac{1}{2}({\bf k}+{\bf k}')$ and assume a linear (or linearized) dependence on ${\bf K}$:
\begin{align}
 {\bf p}^{\rm vc}_{\bf K} = {\bf p}^{\rm vc}_0 + ({\bm \alpha}\cdot {\bf K}){\bf p}^{\rm vc}_1.
\end{align}
For notational convenience, we write ${\bf e}\cdot{\bf p}^{\rm vc}_{\bf K} = p^{\rm vc}_{\bf K} = p^{\rm vc}_0 + ({\bm \alpha}\cdot {\bf K})p^{\rm vc}_1$. The light-matter matrix element is finally written as:
\begin{align}
 h^{\rm v,c}_{{\bf k}',{\bf k}} = h^{\rm vc}_{{\bm \kappa},{\bf K}} = A_{\bm \kappa} p_{\bf K}^{\rm vc}.
\end{align}
This expression makes it immediately clear that the wave vector $\bm \kappa$ is exclusively selected by properties of the light field, whereas the wave vector $\bf K$ is exclusively determined by material properties. In the following, we will find that, in the case of exciton transitions, $\bm \kappa$ ($\bf K$)  is related to the COM (relative) momentum of the exciton. 

\subsection{Exciton transitions}
We are now interested in the transition amplitude for exciton formation ${\cal T}_X \equiv \langle X | H_{\rm LM} | {\rm vac} \rangle$. Here, $| X \rangle$ denotes an excitonic state, which in 2D is characterized through four quantum numbers for relative and COM motion. We choose $|X\rangle = |{\bf k}_{\rm com},n,j\rangle$, i.e. we describe the excitonic state by its linear COM momentum ${\bf k}_{\rm com}$, and its hydrogenic quantum numbers $n$ and $j$, representing the relative degrees of freedom. The vacuum state $|{\rm vac}\rangle$ corresponds to a filled valence band and an empty conduction band.

The excitonic wave function can be written as
\begin{align}
  \langle {\bf R}, {\bf r} | X \rangle \equiv
  \Phi_{n,j}^{{\bf k}_{\rm com}}({\bf R},{\bf r}) \equiv
  \Phi_{{\bf k}_{\rm com}}^{\rm (com)}({\bf R}) \times
  \Phi_{n,j}^{\rm (rel)}({\bf r}),
\end{align}
where ${\bf r}={\bf r}_{\rm e}-{\bf r}_h$ are relative coordinates of an electron-hole pair, and ${\bf R}=\frac{1}{2}({\bf r}_{\rm e}+ {\bf r}_{\rm h})$ are the COM coordinates. The relative motion of electron and hole is described by the solutions to the 2D hydrogen atom, which are given by \cite{haugkoch}
\begin{align}
\label{phirel}
 \Phi_{n,j}^{\rm (rel)}({\bf r}) = \tilde{\cal N}_{n,j} f_{nj}(r) e^{i j \phi_r} = \tilde{\cal N}_{n,j} \rho^{|j|} e^{-\frac{|\rho|}{2}} L_{n-|j|}^{2|j|}(\rho) e^{i j \phi_r},
\end{align}
where $\rho=r\rho_n$, with the inverse length scale given by $\rho_n=\frac{2}{(n+1/2)a_0}$. The material-specific length scale $a_0=\hbar^2 \epsilon/(e^2M)$ is the effective Bohr radius depending on effective mass $M$ and dielectric constant $\epsilon$. The normalization of the relative wave function is given by 
\begin{align}
\tilde{\cal N}_{n,j} = \sqrt{\frac{(n-|j|)!}{(n+|j|)!} \left(\frac{\rho_n}{2}\right)^2 \frac{1}{\pi} \frac{1}{n+1/2}}.
\end{align}
For the COM part, we simply assume plane waves, $\Phi_{{\bf k}_{\rm com}}^{({\rm com})}({\bf R}) = e^{i {\bf k}_{\rm com}\cdot {\bf R}}$.

Without making use of the explicit solution for the excitonic wave functions, we write for the transition amplitude:
\begin{align}
\label{AX}
 {\cal T}_X &=  \int d^2{\bf R} \int d^2{\bf r} \sum_{{\bm \kappa},{\bf K}} \langle X|{\bf R},{\bf r} \rangle \langle {\bf R},{\bf r} | {\bm \kappa},{\bf K} \rangle 
 \times \nonumber \\ & \times \langle {\bm \kappa},{\bf K} | H_{\rm LM} | {\rm vac} \rangle.
\end{align}
The last term corresponds to the band-to-band transition amplitude evaluated above,
\begin{align}
 \langle {\bm \kappa},{\bf K} | H_{\rm LM} | {\rm vac} \rangle &= \langle {\bf K}-\frac{\bm \kappa}{2} | H_{\rm LM} | {\bf K} + \frac{\bm \kappa}{2} \rangle = \nonumber \\ &= 
h^{\rm v,c}_{{\bf K} - \frac{\bm \kappa}{2},{\bf K} + \frac{\bm \kappa}{2}} = A_{\bm \kappa} p_{\bf K}^{\rm vc},
\end{align}
which is Fourier transformed to spatial coordinates by the second term,
\begin{align}
 \langle {\bf R},{\bf r} | {\bm \kappa},{\bf K} \rangle = \frac{1}{S^2} e^{i ({\bf K}+\frac{\bm \kappa}{2}) \cdot{\bf r}_{\rm e} - i ({\bf K}-\frac{\bm \kappa}{2}) \cdot{\bf r}_{\rm h} } = \frac{1}{S^2} e^{i ({\bf K}\cdot{\bf r} + i{\bm \kappa}\cdot{\bf R})}.
\end{align}
This expression explicitly shows that the wave vector ${\bm \kappa}$ ($\bf K$) is conjugate to the COM (relative) variable. 

Plugging all expressions into Eq.~(\ref{AX}), the ${\bf R}$ integral is immediately evaluated into a Kronecker-Delta $\delta_{{\bm \kappa},{\bf k}_{\rm com}}$, so photon momentum ${\bm \kappa}$ and COM momentum ${\bf k}_{\rm com}$ must match.
 We obtain:
\begin{align}
\label{AX2}
 {\cal T}_X &= \frac{(2\pi)^2 \tilde {\cal N}_{nj}}{S^2} A_{{\bf k}_{\rm com}} \int d^2{\bf r} f_{nj}(r) e^{ij \phi_r} \times \nonumber \\  & \times \sum_{\bf K} e^{i {\bf K}\cdot{\bf r}} (p_0^{\rm vc} + {\bm \alpha}\cdot{\bf K} p_1^{\rm vc}).
\end{align}
The ${\bf K}$-sum is a Fourier transform into the relative variable ${\bf r}$, and we can write $\sum_{\bf K} e^{i {\bf K}\cdot{\bf r}} (p_0^{\rm vc} + {\bm \alpha}\cdot{\bf K} p_1^{\rm vc}) = (2\pi)^2 [p^{\rm vc}_0 \delta({\bf r}) - i p_1^{\rm vc} {\bm \alpha}\cdot \nabla_{\bm r} \delta({\bf r})]$. From this expression it can immediately be seen that a constant dipole moment leads to non-zero transition amplitudes only if the relative exciton wavefunction $f_{nj}(r)$ is non-zero at $r=0$. This is the case only for $s$-excitons. The linear dependence of the dipole momentum on ${\bf K}$, expressed by the second term, gives rise to non-vanishing transition amplitudes if the first derivative of $f_{nj}(r)$ is non-zero at $r=0$. 

This second term  enables the formation of excitons in higher momentum states than the $s$ series. However, we emphasize that this term is independent from the light source, and with respect to the relative degrees of freedom ($n,j$), we get the same transitions, no matter what the spatial structure of the light might be (as long as approximation (A1) holds). Our analysis shows that the Elliott formula is unchanged by spatial structure of light beyond the scale of the unit cell.

Our result agrees with a recent experiment in 2D transition metal dichalcogenides (TMDs) where twisted light has been used to reveal light-like exciton dispersion \cite{simbulan20}. Using non-resonant Laguerre-Gaussian beams, they observed a blue shift of the exciton energy that increased with $\ell$; this indicates that the OAM was transferred preferentially to the COM of the exciton during its creation. However, we stress that our results apply much more generally since no particular 2D semiconductor or spatial light profile was specified during the analysis. This implies that the dispersion of all 2D excitons could be probed in a similar manner, presenting an alternative method to the traditional angle-resolved photoluminescence measurements \cite{kwon19}.

\section{Example: Excitons from twisted light}
\begin{figure*}
	\centering
	\includegraphics[scale=0.6]{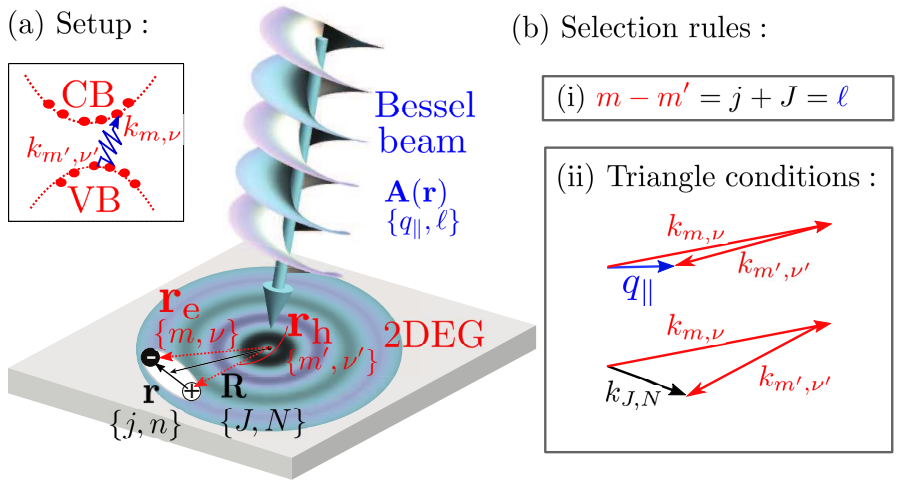}
	\caption{
		(a) A Bessel light beam ${\bf A}$ with photon momentum/angular momentum $(q_{\|},\ell)$ creates an electron-hole pair in a 2D electron gas (2DEG). The electron [hole] is characterized by quantum numbers $(m,\nu)$ [$(m',\nu')$] for angular momentum/ momentum. The pair can form a bound state, and the degrees of freedom of such an exciton are the center-of-mass motion, characterized by angular momentum/ momentum quantum numbers $(J,N)$, and the relative motion, characterized by quantum numbers $(j,n)$ for angular momentum and energy. (b) The selection rules for optical transitions and exciton formation reflect (i) conservation of angular momentum, and (ii) conservation of linear momentum, reflected by the illustrated triangle conditions.} 
	\label{Tnu}
\end{figure*}

Thus far, we have been very general in our treatment with respect to the profile of the optical excitation. In the following, we are going to treat the specific case of a Bessel beam and, besides the analytical treatment along the lines presented in the previous sections, we will also present the result of numerical evaluations. With this choice of the vector potential, our system exhibits a cylindrical symmetry, since the light field has an azimuthal phase dependence $\exp(i \ell \phi)$, where $\ell$ defines the OAM per photon (in units $\hbar$). To match this symmetry, we consider a cylindrical sample, noting that the sample geometry becomes irrelevant in the thermodynamic limit. Accordingly, we adopt our theoretical description to this symmetry, and express the light-matter coupling in terms of a cylindrical wave functions, see also Refs.  \onlinecite{quinteiro10,farias13}. We illustrate this case in Fig.~\ref{Tnu}, where we also sketch the resulting selection rules in terms of the good quantum numbers for the cylindrical symmetry. 

\subsection{Band-to-band transitions in cylindrical basis}
 Instead of plane waves with linear momentum quantum number, the electronic basis in a cylindrical sample is best described by wave functions $ \varphi_{m,\nu}({\bf r}) = {\cal N}_{m,\nu} J_m(k_{m,\nu} r) \exp(i m \phi)$, which solve the Schr\"odinger equation for free electrons with cylindrical boundary conditions. Here, $J_m(x)$ denotes the $m$th Bessel function, and the momenta $k_{m,\nu}$ must be chosen such that the wave function vanishes at the system boundary (i.e. for $|{\bf r}|=R_0$). Therefore, we have $k_{m,\nu}=x_{m,\nu}/R_0$, with $x_{m,\nu}$ the $\nu$th zero of the $m$th Bessel function. The normalization is given by ${\cal N}_{m,\nu} = (R_0\sqrt{\pi} |J_{|m-1|}(x_{m,\nu})|)^{-1}$. As in the previous section, the crystal lattice is taken into account by multiplying the wave functions $\varphi_{m,\nu}$ with lattice-periodic Bloch functions $u_\lambda({\bf r})$ for the bands $\lambda$. For simplicity, the Bloch functions are assumed to be independent from the quantum numbers $m$ and $\nu$. With this, the electronic basis is given
\begin{align}
\varphi_{\lambda;m,\nu}({\bf r})  = {\cal N}_{m,\nu} J_m(k_{m,\nu}r) \exp(im\phi) u_{\lambda}({\bf r}).
\end{align}

In this basis, the light-matter transition amplitudes in the Coulomb gauge (within the weak field limit) are given by
\begin{align}
\label{sumcell}
h^{\rm v,c}_{m,\nu;m',\nu'} =  \frac{-i\hbar e}{SM} \int d{\bf r} \bar \varphi_{+, m',\nu'}({\bf r}) \left[ {\bf A}({\bf r})\cdot \nabla_{\bf r} \right] \varphi_{-, m,\nu}({\bf r}),
\end{align}
with $\bar \varphi$ denoting the complex conjugate of $\varphi$. 

Before we proceed, let us first fix the vector potential. We consider a vector potential which in the sample plane reads  ${\bf A}({\bf R}) = A_0 a(R) e^{i\ell \phi} {\bf e}_{\sigma}$, where ${\bf e}_{\sigma}$ is the polarization in the plane. For a circularly polarized Bessel beam with OAM $\ell$, we have $a(R) = J_\ell(q_{\|} R)$, with $q_{\|}$ the in-plane photon momentum. Note that a vertical contribution to the vector potential, needed to fulfill Maxwell's equation, is neglected here since it is not relevant for light-matter interaction with a two-dimensional medium. However, it is important to keep in mind that the frequency $\omega$ of the photon depends also on the perpendicular momentum component $q_z$, $\omega = \frac{c}{\hbar}\sqrt{q_{\|}^2 +q_z^2}$.

Invoking the approximations (A1) and (A2), both ${\bf A}({\bf r})$ and $J_m(k_{m,\nu}r) \exp(im\phi)$ can be considered constant on the scale of the lattice constant $a$. Thus, the evaluation of $h^{\rm v,c}_{m,\nu;m',\nu'}$ can be split into  an integral $\cal I$ restricted to the unit cell and a sum over units cells ${\cal S}_{m,\nu;m',\nu'}$, that is, we can write $h^{\rm v,c}_{m,\nu;m',\nu'}=   \frac{-i\hbar }{ S M} {\cal S}_{m,\nu;m',\nu'} \times {\cal I}$. As before, the sum over the whole system takes into account the variation of the light field, occurring on larger scales, whereas the unit cell integral determines the material's dipole moment taken to be a constant.

Explicitly, the two contributions are given by
\begin{align}
{\cal I} =  e \int_{\rm c} {\rm d}{\bf r}\bar u_+({\bf r}) ( {\bf e}_\sigma \cdot \nabla_{\bf r} )  u_-({\bf r}),
\end{align}
and
\begin{align}
{\cal S}_{m,\nu;m',\nu'} =&  {\cal N}_{m,\nu}  {\cal N}_{m',\nu'} A_0 \sum_i J_m(k_{m,\nu}R_i)  J_{m'}(k_{m',\nu'}R_i) \nonumber \\ &
 \times  J_\ell(q_{\|} R_i) \exp[i(m-m'+\ell)\phi_i],
\end{align}
where $(R_i,\phi_i)$ denote the lattice vectors. In analogy to Eq.~(\ref{eqn:hvc}), we can read off the results of the  cell integral ${\cal I}$ as the  interband dipole moment $p_{\rm vc} \equiv  {\bf e}_{\sigma} \cdot {\bf d} = {\cal I}$, which depends only on the material. Since we take it to be constant here, it will enter the transition amplitudes only as a prefactor. To evaluate ${\cal S}_{m,\nu;m',\nu'}$, we replace the summation over cells by an integral. With this, we immediately arrive at a first selection rule from the angular part of the integral: 
\begin{align}
\label{kronS}
{\cal S}_{m,\nu;m',\nu'} \propto  \delta_{\ell+m-m'}.
\end{align}
 The radial integral in ${\cal S}_{m,\nu;m',\nu'}$ is over a product of three Bessel functions, ${\cal S}_{m,\nu;m',\nu'} \propto \int_0^{R_0} dr \ r J_m(k_{m,\nu}r)  J_m'(k_{m',\nu'}r) J_\ell(q_{\|} r)$. Its analytic solution (in the limit $R_0 \rightarrow \infty$) has been derived in Ref. \onlinecite{jackson72}, and can also be found in the Supplemental Material (SM). Here, we only consider that, from this solution, the integral takes non-zero values only if a triangle condition is fulfilled: The three scalars $k_{m,\nu}$, $k_{m',\nu'}$, and $q_{\|}$ must be such that they can form a triangle (including the limit in which the triangle is squeezed to a line). Therefore, this condition yields a second selection rule: the change of electron momentum upon a band-to-band transition is bounded by the in-plane momentum of the photon.

\subsection{Exciton transitions}
The amplitudes $h^{\rm v,c}_{m,\nu;m',\nu'}$ quantify the band-to-band transition which generates an electron-hole pair characterized by $m-m'=\ell$ and $|k_{m,\nu}-k_{m',\nu'}| \leq q_{\|}$. Next, we have to ask which excitonic states can be formed from these pairs. In accordance with the presumed cylindrical symmetry of the system, we now also describe the excitonic states in terms of cylindrical-symmetric quantum numbers, $|X\rangle = |N,J,n,j\rangle$, where $n,j$ account for the state of relative motion (as before), and $N,J$ for the COM degrees of freedom (instead of ${\bf k}_{\rm com}$ used in the previous section). 
Again, the excitonic wave function is a product of the relative ($\bf{r}$) and COM ($\bf{R}$) contributions: $\langle {\bf R},{\bf r}|X \rangle \equiv \Phi_{n,j}^{J,N}({\bf R},{\bf r}) \equiv \Phi_{N,J}^{\rm (com)}({\bf R}) \times \Phi_{n,j}^{\rm (rel)}({\bf r})$. The relative part is unchanged, given by Eq.~(\ref{phirel}). Since the COM of the exciton is subject to the same boundary conditions as electron and hole individually, its wave function is given by:
\begin{align}
 \Phi_{J,N}^{\rm (com)}({\bf R}) = {\cal N}_{J,N} J_J(k_{J,N} R) \exp(i J \phi_{\rm com}),
\end{align}
where the quantum number $J$ denotes the angular momentum of the COM, and both $J$ and $N$ together define the total COM momentum $Q_{\rm com} = x_{J,N}/R_0$. 

Projecting the excitonic wave function onto the rotationally symmetric basis for electron and hole wave functions is equivalent to a Hankel transform. This projection yields a quantity 
$ {\cal B}_{m,\nu;m',\nu'}^{n,j;J,N}$:
\begin{align}
\label{B}
 {\cal B}_{m,\nu;m',\nu'}^{n,j;J,N} =& {\cal N}_{m,\nu}  {\cal N}_{m',\nu'} \int {\rm d}{\bf r}_{\rm e} \int {\rm d}{\bf r}_{\rm h}  
  \bar \Phi_{n,j}^{J,N}({\bf R},{\bf r}) \nonumber \\ &
   J_m(k_{m,\nu}r_{\rm e})  J_{m'}(k_{m',\nu'}r_{\rm h})  \exp[i(m \phi_{\rm e} - m' \phi_{\rm h})].
\end{align}
An explicit analytic expression which solves this integral is provided in the SM. As before for ${\cal S}_{m,\nu;m',\nu'}$, we also encounter a triangle condition in the evaluation of $ {\cal B}_{m,\nu;m',\nu'}^{n,j;J,N}$: it is non-zero, only if the lengths $k_{m,\nu}$, $k_{m',\nu'}$, and $k_{J,N}$ form a triangle, i.e. $ |k_{m,\nu}-k_{m',\nu'}|\leq k_{J,N}$. More importantly, as shown in the SM, one of the integrals in Eq. (\ref{B}) yields a Kronecker-$\delta$:
\begin{align}
\label{kronB}
  {\cal B}_{m,\nu;m'\nu'}^{n,j;J,N} \sim \delta_{j+J,m-m'}.
\end{align}
Together with the selection rule for band-to-band transitions, Eq.~(\ref{kronS}), Eq.~(\ref{kronB}) reflects conservation of angular momentum. 

We are now in the position to calculate the exciton transition amplitude ${\cal T}_{n,j}^{J,N} \equiv \langle X_{n,l}^{J,N} | H_{\rm LM} | {\rm vac} \rangle$:
\begin{align}
 {\cal T}_{n,j}^{J,N} = \sum_{m,\nu;m',\nu'} h^{\rm v,c}_{m,\nu;m',\nu'} {\cal B}_{m,\nu;m'\nu'}^{n,j;J,N}.
 \label{Tnj}
\end{align}
These sums should go over all occupied (empty) levels $m',\nu'$ ($m,\nu$), but a more practical limitation of these sums is due to the fact that ${\cal B}_{m,\nu;m'\nu'}^{n,j;J,N} \approx 0$ when either $k_{m,\nu}$ or $k_{m',\nu'}$ become much larger than the inverse of the Bohr radius, $a_0^{-1}$. 

\subsection{Numerical evaluation}
The last observation allows us to introduce a cutoff momentum $k_{\rm cut} \gg a_0^{-1}$ at which the sums can be truncated. With this, the numerical evaluation of Eq.~(\ref{Tnj}) become feasible. For concreteness, by comparison of ${\cal T}_{n,j}^{J,N}$ obtained from different cutoff momenta $k_{\rm cut}$, we estimate that the relative error remains below 0.1 for $k_{\rm cut} a_0 \geq 3.75$. In the numerical evaluation of ${\cal T}_{n,j}^{J,N}$ presented below, we have included 47,100 Bessel functions. With that, $k_{\rm cut} a_0>3.75$ for system sizes up to $R_0/a_0=12,500$. We note that $k_{\rm cut}$ also restricts the sums in $m$ and $\nu$ in the following way: $|m|<\pi k_{\rm cut} R$ and/or $\nu<k_{\rm cut} R$.

The numerical evaluation confirms the analytical result from Sec. II that the Elliott formula remains unchanged by the spatial structure of the light source. To this end, we obtained the height of the spectral lines, $\tilde {\cal T}_{n,j}$, by summing the contributions from all COM momentum modes at a given $n$ and $j$:
\begin{align}
\tilde {\cal T}_{n,j} = \frac{1}{{\cal N}_{\cal T}} \sum_{N,J}  {\cal T}_{n,j}^{J,N}.
\end{align}
To make this quantity independent from the intensity of the light, we normalize by ${\cal N}_{\cal T} = \sqrt{ \sum_{n,j,N}   \left| {\cal T}_{n,j}^{J,N} \right|^2 }$.
The results are shown in Table~\ref{tab:1} for a system of size $R_0=10^4 a_0$ in a Bessel beam with OAM $\ell=0$ and $\ell=1$ and in-plane photon momentum $q_{\|}=10^{-3}a_0^{-1}$. For comparison, we also provide the results from the 2D Elliott formula for an infinite system. All values agree very well with each other.
\begin{table}[t]
    \centering
    \begin{tabular}{|c||c|c|c|}
    \hline
              & 1s & 2s & 3s  \\
   \hline
   \hline
    OAM 0     & 0.998 & 0.037 & 0.0079 \\
    OAM 1     & 0.995 & 0.037 & 0.0079 \\
   2D Elliott & 0.9993 & 0.037 & 0.0080\\
   \hline
    \end{tabular}
    \caption{Relative transition strength, $|\tilde {\cal T}_{n,j}|^2$, for a system of size $R_0=10^4 a_0$ in a Bessel beam with OAM $\ell=0$ and $\ell=1$ and in-plane photon momentum $q_{\|}=10^{-3}a_0^{-1}$. For comparison, we also provide the results from 2D Elliott formula for an infinite system in a Gaussian beam.}
    \label{tab:1}
\end{table}

We note that the numerical evaluation also yields small but finite values for transitions into $p$-states. However, in contrast to the values for the transitions into $s$-states, these values show a strong and non-monotonic dependence on the system size and/or photon momentum. This suggests that, in accordance with our general arguments presented in Sec. II, the finite transition amplitudes into $p$-states are numerical artifacts, and the only bright transitions occur into the $s$ states.

\begin{figure}[t]
	\centering
	\includegraphics[scale=0.25]{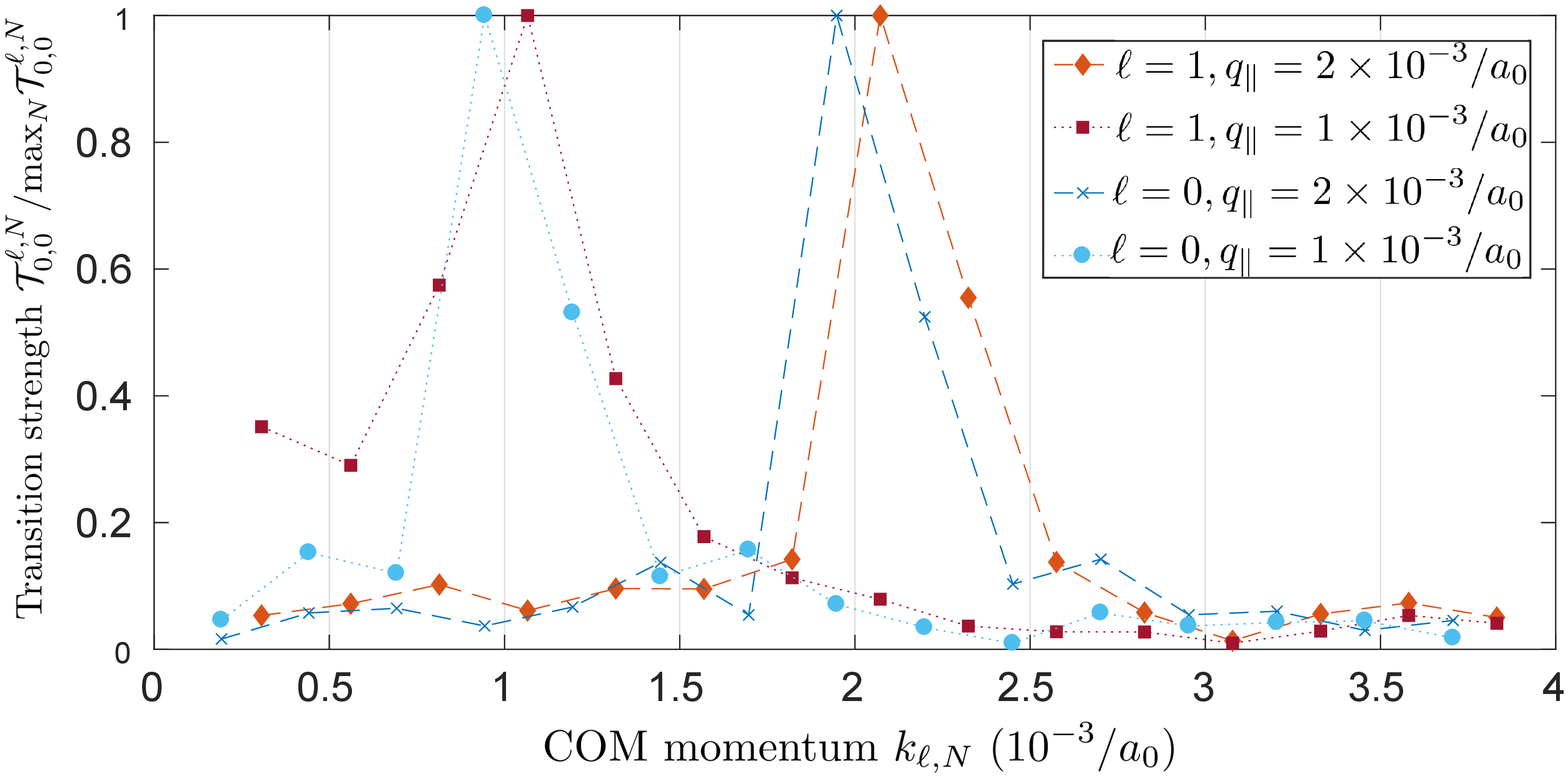}
	\caption{For the 1$s$ transition, we plot the transition strength ${\cal T}_{0,0}^{\ell,N}$ into the different COM modes $k_{\ell,N}$, normalized by the peak value ${\rm max}_N ( {\cal T}_{0,0}^{\ell,N} )$, for illumination with $\ell=0$ and $\ell=1$ Bessel beams. The peak is obtained for the best match between COM momentum $k_{\ell,N}$ and in-plane photon momentum, $q_{\|}=10^{-3}a_0^{-1}$.
	\label{fig:results}}
\end{figure}

Our numerical evaluation also confirms the selection rule that the COM momentum of the exciton is determined by the linear in-plane momentum of the photon. To this end, we focus on the 1$s$ transition and evaluate the transition strengths ${\cal T}_{0,0}^{\ell,N}$ into the different COM modes $k_{\ell,N}$. The results, normalized by the peak value ${\rm max}_N ( {\cal T}_{0,0}^{\ell,N} )$, are shown in Fig. \ref{fig:results} for different values of photon OAM $\ell$ and photon momentum. The transition strength is clearly peaked for the COM momenta which match with the momentum of the photon, but barely depends on the OAM. 

Let us finally discuss the different length scales which appear in the calculation, that is, the effective Bohr radius $a_0$, the sample size $R_0$, and the inverse of the photon momentum $q_{\|}^{-1}$.  In the calculation, we have taken the effective Bohr radius $a_0$ as the unit of length.  
With typical values of the dielectric constant being much greater than 1 (e.g. $\approx 13$ in GaAs\cite{Strazalkowski76} and 7 in semiconducting TMDs\cite{Laturia18}), and the effective mass being much smaller than the electron mass (e.g. $0.067$ and $0.39$ electron masses for the conduction band in GaAs\cite{Cardona61} and model TMDs\cite{Kormanyos15}, respectively), the effective Bohr radius can significantly exceed the Bohr radius of the hydrogen atom ($\approx 0.05$nm). Typical values range between 0.1 to 1 nm. Taking the numerical constraints into account (i.e. truncation errors), our study examines sample sizes $R_0$ on the order of $10^4$ effective Bohr radii which corresponds to sample sizes on the order of 1-10 microns. Importantly, this size is significantly larger than the optical vortex. Regarding the in-plane photon momentum $q_{\|}$, an upper limit is given by
the inverse of the wavelength, $2\pi/\lambda_0$, assuming vertical incidence on the sample. The wave length $\lambda_0$ is determined by the band gap of the material. As an estimate for this limit, we obtain $100{\rm nm}^{-1}$. Thus, our choice of $q_{\|}=10^{-3}a_0^{-1}$ corresponds to the upper limit if $a_0=0.1$ nm, while this choice remains below that limit if $a_0$ is larger.

\section{Summary and Conclusions} 

We have shown that the vector potential selects the COM quantum numbers (absolute value of COM momentum + COM angular momentum), but has no effect on the transition amplitudes into states with different relative quantum numbers $n$ and $j$. This implies that Elliott's formula is unchanged by the structure in the light field. The approximation which gives rise to these conclusion is the separation of length scales: unit cell vs. wavelength. This assumption implies that $A(x)$ and $e^{iqx}$ are constant on the the level of a unit cell, and for the dipole moment, $p_{{\bf k},{\bf k}+{\bf q}}^{\rm vc} \approx p_{{\bf k},{\bf k}}^{\rm vc}$. We have confirmed our general analytical result by performing numerical evaluations for the concrete case of Bessel beams in a circularly symmetric sample. Qualitatively, we have shown that, for a transition to be optically bright, the sum of both relative and COM angular momenta, $j+J$, must be equal to the OAM value $\ell$ of the light. Quantitatively, we have evaluated that the transition amplitudes are given by the Elliott formula. 

While our results rule out twisted light for the generation of dark excitons, the predicted transfer of OAM to the COM degree of freedom can be useful from the perspective of quantum simulation, and especially from the point of view of artificial gauge fields. In Ref.~\cite{joerg20}, it has been shown that artificial flux is generated in a photonic system when OAM light is injected into a waveguide lattice. Excitons in tunable lattices have recently be shown to form strongly correlated many-body phases, such as Mott insulating phases \cite{lagoin21} or checkerboard phases \cite{lagoin22}. If, in the future, twisted light provided excitonic lattices with artificial magnetic fluxes, this could give rise to chiral Mott insulators \cite{dhar12} or extended supersolid regimes \cite{suthar20}.

\acknowledgments{
We thank Bin Cao, Valentin Kasper, Maciej Lewenstein, Andrey Grankin, Sunil Mittal, Jay Sau, Deric Session, Glenn Solomon, Daniel Suarez, and Jon Vanucci for thoughtful comments and fruitful discussions.
T.G. and U.B. acknowledge funding from “la Caixa” Foundation (ID 100010434, fellowship code LCF/BQ/PI19/11690013), the European Research Council for ERC Advanced Grant NOQIA; Agencia Estatal de
Investigación (R\&D project CEX2019-000910-S, funded by 
MCIN/AEI/10.13039/501100011033, Plan National FIDEUA PID2019-106901GB-I00, FPI, QUANTERA MAQS PCI2019-111828-2, Proyectos de I+D+I “Retos Colaboración”
RTC2019-007196-7); Fundació Cellex; Fundació Mir-Puig; Generalitat de Catalunya through the CERCA program, AGAUR Grant No. 2017 SGR 134, QuantumCAT  U16-011424, co-funded by ERDF Operational Program of Catalonia 2014-2020; EU Horizon 2020 FET-OPEN OPTOLogic (Grant No 899794); National Science Centre, Poland (Symfonia Grant No. 2016/20/W/ST4/00314); Marie Sk\l odowska-Curie grant STREDCH No 101029393; “La Caixa” Junior Leaders fellowships (ID100010434) and EU Horizon 2020 under Marie Sk\l odowska-Curie grant agreement No 847648 (LCF/BQ/PI20/11760031, LCF/BQ/PR20/11770012, LCF/BQ/PR21/11840013).
J.S. acknowledges support from the NSF Graduate Research Fellowship Program (GRFP) and the ARCS Foundation. J.S. and M.H. were supported by AFOSR FA95502010223, NSF PHY1820938, and NSF DMR-2019444, ARL W911NF1920181, ARO W911NF2010232, Simons and Minta Martin Foundations.  
}

\onecolumngrid
\appendix
\section{Band-to-band transition matrix for a Bessel beam}
The band wave functions are expressed in terms of Bessel functions $J_m(k_{m,\nu} r)$, and the spatial profile of the light is given by a Bessel beam $ J_\ell(q_{\|} r)$. Thus, the matrix elements for  band-to-band transitions are proportional to the integral over a product of three Bessel functions:
\begin{align}
 {\cal S}_{m,\nu;m',\nu'}  \propto  \int_0^{R_0} dr \ r J_m(k_{m,\nu}r)  J_m'(k_{m',\nu'}r) J_\ell(q_{\|} r).
\end{align}
To evaluate this integral, we take the limit $R_0 \rightarrow \infty$, and follow the procedure as described in Ref.~\onlinecite{jackson72};  that is, we perform a plane-wave expansion of the Bessel functions. With this, the integral is found to take the following value:

\begin{align}
 {\cal J} & \equiv \int_0^{\infty} dr \ r J_m(k_{m,\nu}r)  J_m'(k_{m',\nu'}r) J_\ell(q_{\|} r)   =
 \begin{cases}
 \frac{1}{2\pi A_{\triangle}} \cos \left( m \alpha_2 - m' \alpha_1 \right) ,& \text{if } k_{m,\nu},k_{m',\nu'}, q_{\|} \text{can form  a triangle } , \\
 0  & \text{otherwise.}
\end{cases}
\end{align}
If a triangle with lengths $ k_{m,\nu},k_{m',\nu'}$, and  $q_{\|}$ can be formed, the quantity $A_{\triangle}$ denotes the area of this triangle, and $\alpha_1$ and $\alpha_2$ are exterior angles of this triangle. Defining $\kappa \equiv (k_1+k_2+k_3)/2$, we have
\begin{align}
 A_{\triangle} = \sqrt{\kappa (\kappa-k_1)(\kappa-k_2)(\kappa-k_3)}.
\end{align}
The angles are given by:
\begin{align}
 \alpha_1 &= {\rm arccos}\left( \frac{k_1^2-k_2^2-k_3^2}{2k_2k_3} \right), \\
 \alpha_2 &= {\rm arccos}\left( \frac{k_2^2-k_3^2-k_1^2}{2k_1k_3} \right).
\end{align}

\section{Hankel transform of the exciton}
In a circularly-symmetric system, the electronic bands are conveniently given by Bessel functions. Decomposing the excitonic wave function in terms of these single-particle states is equivalent to a Hankel transform of the exciton. Let the relative motion of electron and hole of the exciton be described by a principal quantum number $n$, and an angular momentum quantum number $j$, and the center-of-mass (COM)motion be described by quantum numbers $J$ for angular momentum and momentum and $N$ for momentum/energy. The overlap of such exciton with an electron in a state described by quantum numbers $m,\nu$, and a hole described by $m',\nu'$ reads:
\begin{align}
 {\cal B}_{m,\nu;m',\nu'}^{n,j;J,N} =  {\cal N}_{m,\nu} {\cal N}_{m',\nu'} \int{\rm d}{\bf r}_{\rm e} \int{\rm d}{\bf r}_{\rm h} \bar \Phi_{J,N}^{\rm (com)} ({\bf R}) \bar \Phi_{n,j}^{\rm (rel)} ({\bf r}) J_m(k_{m,\nu} r_{\rm e}) J_{m'}(k_{m',\nu'} r_{\rm h}) \exp\left[ i (m \phi_{\rm e} - m' \phi_{\rm h}) \right]. 
\end{align}
Here, ${\bf R} = ({\bf r}_{\rm e} + {\bf r}_{\rm h})/2$ describes the COM motion, and ${\bf r} = {\bf r}_{\rm e} - {\bf r}_{\rm h}$ describes the relative motion, and ${\bf r}_{\rm e,h}$ are expressed in polar coordinates $(r_{\rm e,h}, \phi_{\rm e,h})$. As a first step to evaluate $\cal B$, we will expand the Bessel functions (including the one contained in the definition of $\bar \Phi_{J,N}^{\rm (com)}$, see main text) in terms of plane waves. To this end, we note that 
\begin{equation}
    e^{iz\cos(\theta)}=\sum_{n=-\infty}^{n=\infty}i^n e^{in\theta}J_n(z).
\end{equation}
Therefore,
\begin{equation}
    e^{i{\vec k}\cdot{\vec r}_e}=e^{ikr_e\cos(\phi_e-\phi_k)}=\sum_{n=-\infty}^{n=\infty}i^n e^{in\phi_e}e^{-in\phi_k}J_n(kr_e).
\end{equation}
Integrating both sides over $\int_0^{2\pi}d\phi_k~e^{im\phi_k}$ gives
\begin{equation}
    J_m(kr_e)e^{im\phi_e}=\frac{1}{2\pi}(-i)^m\int_0^{2\pi}e^{i{\vec k}\cdot{\vec r}_e}e^{im\phi_k}d\phi_k.
\end{equation}
Applying this expansion, below we associate ${\bf k}\equiv (k,\phi_k)$ with electron momentum, and ${\bf k'}\equiv (k',\phi_{k'})$ with hole momentum, where $k$ and $k'$ are introduced as short-hand notations for $k_{m,\nu}$ and $k_{m;,\nu'}$. Similarly, the center-of-mass momentum will be associated with a vector ${\bf Q}={Q,\phi_Q}$. With this, and re-expressing electron and hole coordinates in terms of relative and center-of-mass coordinates, we get
\begin{align}
 {\cal B}_{m,\nu;m',\nu'}^{n,j;J,N} = &
 \tilde {\cal N}_{n,j} {\cal N}_{J,N} {\cal N}_{m,\nu} {\cal N}_{m',\nu'} 
 \frac{-i^{m-m'+J}}{(2\pi)^3} \int_0^{2\pi} {\rm d}\phi_k  \int_0^{2\pi} {\rm d}\phi_{k'}   \int_0^{2\pi} {\rm d}\phi_Q 
 \int{\rm d}{\bf r} \int{\rm d}{\bf R}
 e^{i{\bf R}\cdot ({\bf k}-{\bf k}'-{\bf Q})} 
 \times \nonumber \\ & \times
 e^{i{\bf r} \cdot ({\bf k}+{\bf k}')/2} e^{im\phi_k} e^{-im'\phi_{k'}}e^{-iJ\phi_Q}
\rho^{|j|} e^{-\frac{|\rho|}{2}} L_{n-|j|}^{2|j|}(\rho) e^{-i j \phi_{\rm rel}},
 \end{align}
where $\rho \equiv r \rho_n$ with $\rho_n \equiv 2 r/[a_0(n+\frac{1}{2})]$. The integration in ${\bf R}$ yields a $\delta$-function, which imposes a triangle condition for the momenta:
\begin{align}
 \int{\rm d}{\bf R} 
 e^{i{\bf R}\cdot ({\bf k}-{\bf k}'-{\bf Q})}  = (2\pi)^2 \delta^{(2)}({\bf k}-{\bf k}'-{\bf Q}).
\end{align}
Let us next carry out the angular part of the integral in ${\bf r}$.:
\begin{align}
 \int {\rm d}\phi_{\rm rel} e^{-ij \phi_{\rm rel}} e^{i r q \cos(\phi_{\rm rel}-\phi_q)} = 2\pi i^j e^{-ij \phi_q}J_j(qr).
\end{align}
Here, as a short-hand notation, we have introduced ${\bf q}=(q,\phi_q)\equiv \frac{{\bf k}+{\bf k}'}{2}$.  For the radial part of the relative position integral we write:
\begin{align}
f(q,n,j) &  \equiv \int {\rm d}\rho \rho^{j+1}e^{-\frac{\rho}{2}}L^{2j}_{n-j}(\rho)J_j\left(\frac{q}{\rho_n}\rho\right)  = 
\sum_{s=0}^{n-j} \frac{\sqrt{2}(-1)^s(\frac{q}{\rho_n})^j}{\left(\frac{2q^2}{\rho_n^2}+\frac{1}{2}\right)^{j+s+\frac{3}{2}}}
\frac{\Gamma(n+j+1)\Gamma(2j+s+2)}{\Gamma(s+1)\Gamma(j+1) \Gamma(n-j-s+1)\Gamma(2j+s+1)} 
\times \nonumber \\ &
\times~{}_2F_1 \left(\frac{-s}{2},\frac{-1-s}{2},1+j,-\frac{4q^2}{\rho_n^2}\right),
\end{align}
with $~{}_2F_1$ being the hypergeometric function
Note that here we have changed the integration variable from $r$ to $\rho$, which yields a factor $1/\rho_n^2$. For the analytic solution of the integral, we have taken the integration boundary to be at infinity.

Putting all together, we arrive at the following intermediate result:
\begin{align}
\label{eq14}
 {\cal B}_{m,\nu;m',\nu'}^{n,j;J,N} = &
 \tilde {\cal N}_{n,j} {\cal N}_{J,N} {\cal N}_{m,\nu} {\cal N}_{m',\nu'} 
 \frac{i^{m-m'+J+j}}{\rho_n^2} f(q,n,j) \times \nonumber \\ & \times \int_0^{2\pi} {\rm d}\phi_k  \int_0^{2\pi} {\rm d}\phi_{k'} \int_0^{2\pi} {\rm d}\phi_Q
 \delta^{(2)}({\bf k}-{\bf k}'-{\bf Q}) e^{i(m\phi_k - m'\phi_{k'} - j\phi_q -J\phi_Q)}.
\end{align}
Let us for a moment assume $\phi_Q$ to be fixed. Then, in the remaining two integrals, $\phi_k$ and $\phi_{k'}$ will be fixed such that the triangle condition expressed by the $\delta$-function is met.  To proceed, we write the delta function $\delta^{(2)}\left({\vec k}-{\vec k}^\prime-{\vec Q}\right)$ explicitly in terms of $\phi_k$, $\phi_{k^\prime}$, and $\phi_Q$:
\begin{eqnarray}
   \delta^{(2)}\left({\vec k}-{\vec k}^\prime-{\vec Q}\right)=&\delta\left(k\cos\phi_k-k^\prime\cos\phi_{k^\prime}-Q\cos\phi_Q\right)\delta\left(k\sin\phi_k-k^\prime\sin\phi_{k^\prime}-Q\sin\phi_Q\right).
   \end{eqnarray}
To perform the integral over these $\delta$ functions, we first note the identity
\begin{equation}
\label{intid}
    \int d\phi g(\phi)\delta(h(\phi))=\sum_w\frac{g(\phi^w)}{\left|\frac{{\rm d}h(\phi)}{\rm \phi}\right|}\Bigg\rvert_{\phi=\phi^w},
\end{equation}
where $\phi=\phi^w$ denote the solutions to $h(\phi)=0$. We define $h_1(\phi_k) = k\cos\phi_k-k^\prime\cos\phi_{k^\prime}-Q\cos\phi_Q$ and $h_2(\phi_{k'}) = k\sin\phi_k-k^\prime\sin\phi_{k^\prime}-Q\sin\phi_Q$. The zeros of $h_1$ and $h_2$ are given by
\begin{eqnarray}\label{eq_phik}
    \cos\phi_k          &=& \frac{k^\prime\cos\phi_{k^\prime}+Q\cos\phi_Q}{k}, \\
    \sin\phi_{k^\prime} &=& \frac{k       \sin\phi_k         -Q\sin\phi_Q}{k^\prime}.
\end{eqnarray}
If there is a triangle with lengths given by $k$, $k'$, and $Q$, the angles $\phi_k$ and $\phi_{k'}$ of such triangle solve these equations. Note that a second solution can then be obtained from a mirror transformation of the triangle, but since the solutions are equivalent, considering only one of them is sufficient. On the other hand, if such triangle does not exist, the equations (\ref{eq_phik}) have no solution, and the integral is zero.
Explicitly, the solutions can be written as
\begin{eqnarray}
\label{tildephik}
 \phi_k &=& \tilde \phi_k + \phi_Q \ {\rm with} \ \tilde \phi_k = {\rm arccos}\left( \frac{k^2 +Q^2-k'^2}{2kQ} \right), \\
 \phi_{k'} &=& \tilde \phi_{k'} + \phi_Q \ {\rm with} \ \tilde \phi_{k'} = {\rm arccos}\left( \frac{k^2 -Q^2-k'^2}{2k'Q} \right).
\end{eqnarray}
The angle $\phi_q$ in Eq. (\ref{eq14}) is given by:
\begin{eqnarray}
   \label{tildephiq}
   \phi_q&=&\tan^{-1}\left(\frac{q_y}{q_x}\right)=\tan^{-1}\left(\frac{k_y+k^\prime_y}{k_x+k^\prime_x}\right)=\tan^{-1}\left(\frac{k\sin(\tilde\phi_k-\phi_Q)+k^\prime\sin(\tilde\phi_{k^\prime}-\phi_Q)}{k\cos(\tilde\phi_k-\phi_Q)+k^\prime\cos(\tilde\phi_{k^\prime}-\phi_Q)}\right) = \nonumber \\ 
   &=& \tan^{-1}\left(\frac{k\sin(\tilde\phi_k)+k^\prime\sin(\tilde\phi_{k^\prime})}{k\cos(\tilde\phi_k)+k^\prime\cos(\tilde\phi_{k^\prime})}\right)+\phi_Q \equiv \tilde \phi_q +\phi_Q.
\end{eqnarray}
We can also express $q$ in terms of $k$, $k'$, $\tilde \phi_k$, and $\tilde \phi_{k'}$:
\begin{align}
\label{q}
 q=\frac{1}{2}\sqrt{k^2+k'^2 +k k' \cos(\tilde \phi_k - \tilde \phi_{k'}) }.
\end{align}
The term $\left|\frac{{\rm d}h_1(\phi_k)}{{\rm d}\phi_k}\right|^{-1} \times \left|\frac{{\rm d}h_2(\phi_{k'})}{{\rm d}\phi_{k'}}\right|^{-1}$, evaluated at $\phi_k=\tilde \phi_k + \phi_Q$ and $\phi_{k'}=\tilde \phi_{k'} + \phi_Q$, which we get according to Eq. (\ref{intid}) from the two integrals in $\phi_k$ and $\phi_{k'}$, is given by:
\begin{align}
 |h_1'(\tilde \phi_k + \phi_Q)|^{-1} |h_2'(\tilde \phi_k+\phi_Q)|^{-1} = \frac{1}{k k' |\sin(\tilde\phi_k-\tilde\phi_{k'})|}.
\end{align}
Putting all together, we arrive at
\begin{align}
 {\cal B}_{m,\nu;m',\nu'}^{n,j;J,N} = &
 \tilde {\cal N}_{n,j} {\cal N}_{J,N} {\cal N}_{m,\nu} {\cal N}_{m',\nu'} 
 \frac{i^{m-m'+J+j}}{\rho_n^2} f(q,n,j) \frac{1}{k k' |\sin(\tilde\phi_k-\tilde\phi_{k'})|} e^{i (m\tilde \phi_k - m'\tilde \phi_{k'} - j\tilde\phi_q)} 
 \int_0^{2\pi} {\rm d}\phi_Q e^{i(m - m' - j -J)\phi_Q }.
\end{align}
From this, we finally obtain the selection rule $\delta_{m-m',j+J}$ which expresses the conservation of angular momentum. Recalling that $k=k_{m,\nu}$, $k'=k_{m',\nu'}$, $Q=k_{J,N}$, we write as the final result
\begin{align}
 {\cal B}_{m,\nu;m',\nu'}^{n,j;J,N} = \delta_{m-m',j+J}  \tilde {\cal N}_{n,j} {\cal N}_{J,N} {\cal N}_{m,\nu} {\cal N}_{m',\nu'} 
 \frac{2\pi}{\rho_n^2} f(q,n,j) \frac{1}{k_{m,\nu} k_{m',\nu'} |\sin(\tilde\phi_k-\tilde\phi_{k'})|} e^{i (m\tilde \phi_k - m'\tilde \phi_{k'} - j\tilde\phi_q)}.
\end{align}
The angles $\phi_k$, $\phi_{k'}$, $\tilde \phi_q$ are defined in Eqs.~(\ref{tildephik}) and (\ref{tildephiq}). The value of $q$ is given by Eq.~(\ref{q}).

\twocolumngrid

\bibliography{bib2}

\end{document}